\documentstyle[aaspp4,twoside]{article}
\begin{document}

\title{A Planet Orbiting the Star Rho Coronae Borealis}

\author{Robert W. Noyes\footnote{rnoyes@cfa.harvard.edu}, 
 Saurabh Jha\footnote{sjha@cfa.harvard.edu}, 
 Sylvain G. Korzennik\footnote{skorzennik@cfa.harvard.edu},\\
 Martin Krockenberger\footnote{mkrockenberger@cfa.harvard.edu}, and 
 Peter Nisenson\footnote{pnisenson@cfa.harvard.edu}}
\affil{Harvard-Smithsonian Center for Astrophysics, 
60 Garden St, Cambridge, MA 02138}

\author{Timothy M. Brown\footnote{brown@hao.ucar.edu} and 
Edward J. Kennelly\footnote{kennelly@hao.ucar.edu}}
\affil{High Altitude Observatory, National Center for Atmospheric
Research,\\
P.O. Box 3000, Boulder, CO 80307}

\and
\author{Scott D. Horner\footnote{horner@astro.psu.edu}}
\affil{Department of Astronomy and Astrophysics,  Pennsylvania State 
University,\\ 525 Davey Lab, University Park, PA 16802
}

\begin{abstract}
We report the discovery of near-sinusoidal radial velocity variations
of the G0V star $\rho$~CrB, with period 39.6 days and amplitude 67 m
s$^{-1}$.  These variations are consistent with the existence of an
orbital companion in a circular orbit.  Adopting a mass of 1.0
$M_{\odot}$ for the primary, the companion has minimum mass about 1.1
Jupiter masses, and orbital radius about 0.23 AU.  Such an orbital radius
is too large for tidal circularization of an initially eccentric orbit
during the lifetime of the star, and hence we suggest that the low
eccentricity is primordial, as would be expected for a planet formed
in a dissipative circumstellar disk.
\end{abstract}


\section{INTRODUCTION}

Within the past 18 months, our knowledge of low  mass
companions to stars has exploded, so that it is becoming possible to
carry out comparative studies of their properties such as mass,
orbital period, and orbital eccentricity.  This information is opening
the door to a better understanding of the formation and evolution of
planets and brown dwarfs, including the planets in our own solar system.
However, the small number of objects with known orbital parameters does not
yet permit creating a reliable picture of the origin and evolution of such
objects. 

For this reason it is extremely important to increase the number of
known low-mass stellar companions and also gain information on their
orbital properties.  We have developed the Advanced Fiber Optic Echelle (AFOE)
spectrograph in part for this purpose.  Using this instrument we have
begun an intensive program of monitoring about 100 solar-type stars
brighter than V $\sim$ 6.5 with a precision of order 10 m s$^{-1}$;
this is sufficient to detect the reflex orbital velocities of any of
the substellar companions detected by precise radial velocity
techniques to date.  In this Letter we report the detection of
periodic radial velocity variations in the star $\rho$~Coronae
Borealis, which we interpret as due to a Jupiter-mass planet in a near-circular
orbit about the star.

\section{ RHO CORONAE BOREALIS---AN OLD SOLAR-TYPE STAR}

In order to understand the properties of an orbiting companion to
$\rho$~CrB (HD 143761, HR 5968), it is necessary to estimate its mass,
effective temperature, and age. The evidence argues for $\rho$~CrB
being an old, solar type star, with mass close to that of the Sun.

The spectral type of $\rho$~CrB is reported to be either G0V
\markcite{jaschek}(Jaschek, Conde, \& de Sierra 1964) or G2V
\markcite{roman1}\markcite{roman2}(Roman 1955, Roman 1952). Reported
$B-V$ color indices average about 0.61
\markcite{roman2}\markcite{naur}\markcite{argue}(Roman 1955, Naur
1955, Argue 1963), i.e.\ slightly bluer than the Sun, for which
reported values of $B-V$ range from 0.63 to 0.66
(e.g. \markcite{gray1}Gray 1992, \markcite{cayrel}Cayrel de Strobel
1996, \markcite{taylor}Taylor 1994).  The effective temperature
T$_{\rm eff}$ is about 5760K as determined from the infrared flux method
\markcite{gratton}(Gratton, Carretta, \& Castelli 1996 and
\markcite{alonso}Alonso, Arribas, \& Martinez-Roger 1996), although
Gray \markcite{gray2}(1994) concludes from line-depth ratios that
T$_{\rm eff}$ = 5868 K.

The luminosity of $\rho$~CrB is larger than the Sun's, suggesting that
$\rho$~CrB is more evolved and thus older.  This is inferred as
follows. Its V band apparent magnitude is close to V = 5.41
\markcite{rufener}\markcite{naur}\markcite{argue}\markcite{roman2}(Rufener
1976, Naur 1955, Argue 1963, and Roman 1955). A recent (pre-Hipparcos)
estimate of its parallax \markcite{gliese}(Gliese \& Jahreiss 1991) is
$\pi$ = 60$\pm$6 mas, giving a distance of 16.7$\pm$1.7 pc.  These
values imply an absolute visual magnitude M$_V$ = 4.3. Using a
Sun-like bolometric correction of B.C.\ = $-0.07$, this gives a
luminosity of L = 1.61 L$_{\odot}$. Adopting an effective temperature
equal to that of the Sun (5777 K, \markcite{cayrel}Cayrel de Strobel
1996), we obtain a radius of R = 1.27 R$_{\odot}$.

Based on evolutionary H-R diagrams of Bressan et al.\
\markcite{bressan}(1993), the color and luminosity of $\rho$~CrB are
consistent with a 1.0 solar mass star about 10 Gyr old.  The surface
gravity was determined by K\"unzl et al.\ \markcite{kunzli}(1997) and
Gratton et al.\ \markcite{gratton}(1996) to be log(g) = 4.23 and 4.11
respectively, which is also consistent within the uncertainties of the
fitting procedure with an evolved solar mass star.

Henry \markcite{henry}(1997) reports that four years of precise
photometric data for $\rho$~CrB show no significant periodicities
between 1 and 50 days, and constant seasonal means to within 0.00017
magnitudes.  The implied absence of spot activity is consistent with an
old age for $\rho$~CrB.  

The relative chromospheric inactivity of $\rho$~CrB also supports its
high age.  Its Ca II chromospheric flux is below that of the Sun and
shows no significant seasonal variations \markcite{baliunas1}(Baliunas
et al.\ 1995).  In spite of the smaller Ca II flux, the rotation
period of $\rho$~CrB as predicted from that flux is about 20 days
\markcite{noyes1}\markcite{soderblom}(Noyes et al.\ 1984, Soderblom
1985), which is slightly shorter than the 25.4 day sidereal rotation
period of the Sun.  The faster predicted rotation results from the
fact that $\rho$~CrB has slightly bluer $B-V$ color index than the Sun.
There is some observational evidence for a rotational modulation of
the Ca II flux with about a 17-day period
\markcite{baliunas2}(Baliunas et al.\ 1996), although that evidence is
marginal \markcite{soon}(Soon 1997).

Further evidence that $\rho$~CrB is older and more evolved than the
Sun is provided by the metal deficiency; reported values of [Fe/H]
average to $-0.19$\markcite{kunzli}\markcite{hearnshaw}\markcite{wallerstein} 
\markcite{alexander}(K\"unzl et al. 1997,
Hearnshaw 1974, Wallerstein 1962, Alexander 1967).  The ultraviolet
excess $\delta(U-B)$ = 0.06 \markcite{hearnshaw}(Hearnshaw 1974) also
indicates a slightly less than solar abundance.  In addition, $\rho$
CrB is a high proper motion star with a motion out of the galactic
plane of 28 km s$^{-1}$ \markcite{cayrel}(Cayrel de Strobel 1996),
indicating that it may be a member of the old disk population.

The above information,  taken all together, suggests that $\rho$~CrB
is probably of near solar mass but older, near the end of its main
sequence lifetime or possibly already starting hydrogen shell burning
\markcite{cayrel}(cf.\ Cayrel de Strobel 1996, figure 14).  For the
purposes of this Letter, we shall adopt a nominal mass for $\rho$~CrB
of 1.0 M$_{\odot}$.

\section{OBSERVATIONS}

\subsection{Instrumentation and Reduction Procedure}

Data on $\rho$~CrB were taken with the Advanced Fiber-Optic Echelle
(AFOE), located at the 1.5-m telescope of the Whipple Observatory.
This instrument is a cross-dispersed, fiber-fed echelle spectrograph
designed specifically to perform precise stellar radial velocity
measurements \markcite{brown}(Brown et al. 1994); its construction and
operation are a joint project of the Smithsonian Astrophysical
Observatory, the High Altitude Observatory, and Pennsylvania State
University.  The spectrograph covers a wavelength range of about 150
nm between wavelengths of 392 nm and 665 nm, at a typical resolution
$R = \lambda / \delta \lambda$ of 50000.  The combination of a fiber
feed, mechanical stability, and careful thermal control minimizes
drifts in the velocity zero point and also results in a very stable
instrumental resolution profile.  For these observations, wavelength
calibration was provided by an absorption cell filled with iodine
vapor \markcite{horner}(Horner 1996), which superposes a dense
spectrum of molecular iodine (I$_2$) lines upon the stellar spectrum
in the wavelength range between 500 nm and 620 nm.  When operated in
this mode, the spectrograph can achieve Doppler precision of better
than 10 m s$^{-1}$ on a 5.0-magnitude G star in an observation time of
10 minutes \markcite{noyes2}(Noyes et al. 1997).

The method of data reduction is generally similar to that described by
Butler et al.\ \markcite{butler2}(1996), although it differs in
details.  1-D spectra are extracted from the 2-D echelle images,
corrected for scattered light, and divided by the flat-field spectrum
of a continuum lamp.  For each of six spectral orders containing
strong I$_2$ lines the stellar radial velocity is determined
independently, by modeling the observed star-plus-iodine spectrum as
the product of a Doppler-shifted, high signal-to-noise-ratio reference
spectrum of the star alone and a very high spectral resolution
spectrum of the I$_2$ absorption cell in the laboratory reference
frame; the latter was previously measured at the McMath-Pierce FTS
spectrograph at the National Solar Observatory.  The model incorporates
the sought-for relative wavelength shift between the spectrum of the
star and that of the iodine cell, detector shifts due to instabilities
within the spectrograph, and a parametrized wavelength solution and
instrumental resolution profile. The instrumental profile shape is
allowed to vary slowly along the order.

All of the parameters describing this model are adjusted so as to
minimize the rms difference between the observed and modeled spectrum.
The six independently deduced Doppler velocity shifts for each order
are then combined with equal weights to provide the final velocity
shift for that exposure, while the internal scatter amongst the
different orders gives an estimate of its uncertainty. Each
observation normally consists of three consecutive exposures, in order
to correct for cosmic ray contamination; typical exposures have
continuum signal-to-noise ratio per pixel of order 100 to 150.  The
results of the velocity estimates for the three exposures are
corrected for motion of the telescope relative to the solar system
barycenter, and then combined to provide the final measure of the
stellar radial velocity shift and an estimate of its uncertainty.

\subsection{Observations and Results}

Observations of $\rho$~CrB were obtained on 41 nights between 1996 May
and 1997 March.  A tabulation of the data, showing times of
observation, derived velocities, and estimated uncertainties, may be
found on the World Wide Web at {\tt
ftp://cfa-ftp.harvard.edu/pub/afoe/rhoCrB/}.  The data, together with
uncertainties, are plotted in Figure 1.

\begin{figure}[ht]
\plotone{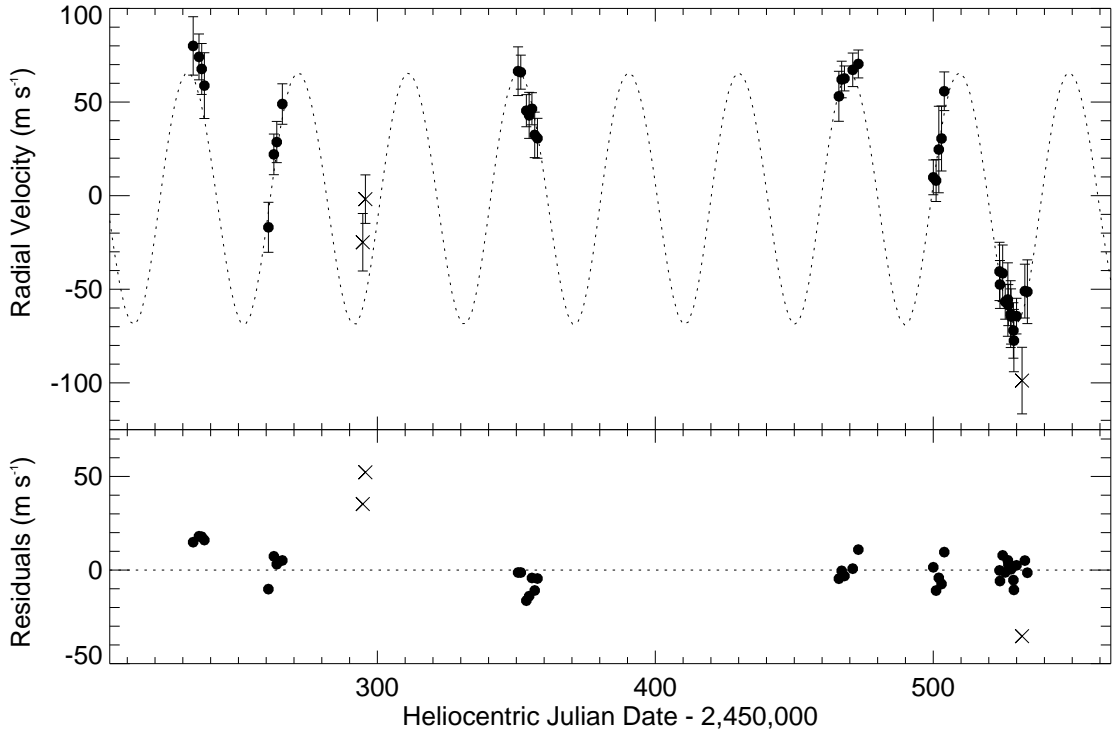}
\caption[]{Top: Radial velocity of $\rho$~CrB (solid
circles), with three 2.5-$\sigma$ rejected points shown as crosses.
Dashed line is an orbital solution using the parameters listed in
Table 1.  Bottom: residuals from the orbital solution. \label{figrv}}
\end{figure}

We fitted a Keplerian orbit to the data using a least-squares fitting
process. The orbital parameters are given in Table 1. They result from
an iterative 2.5-$\sigma$ rejection fitting procedure (where $\sigma$
is the rms of the residuals to the fit); three points were rejected on
this basis, as shown by crosses in Figure 1. We note that the orbital
parameters resulting from a fit without any rejection, or from a
3-$\sigma$ rejection procedure which eliminated only one point, differ
from the values listed in Table 1 by less than their formal
uncertainty. We verified that the elements we derived are the ones
providing the best fit to the data by using a $\chi^2$-minimization
routine based on a genetic algorithm
\markcite{charbonneau}(Charbonneau 1995).  This algorithm uses a form
of directed Monte Carlo search that is specifically designed to locate
the global minimum, regardless of where it lies in the parameter
space. The fit to the next most likely period (at 76.8 days) has
residuals with reduced $\chi^2$ of 3.3, 
corresponding to an extremely small probability that this is the correct
period. 

\begin{deluxetable}{ll}
\tablewidth{4in}
\tablehead{}
\tablecaption{Orbital Parameters for $\rho$~CrB \label{tabparm}}
\startdata
$P$ 		& $39.645 \pm 0.088$ days				\nl
$K_1$ 		& $67.4 \pm 2.2$ $\rm{m\;s^{-1}}$			\nl
$e$		& $0.028 \pm 0.040$ 					\nl
$\omega$ 	& $210 \pm 74$ deg					\nl
$T_{\rm periastron}$& $2,450,413.7 \pm 8.2$ HJD				\nl
									\nl
$a_1 \sin i$	& $(36.75 \pm 0.92) \times 10^{6}$ m			\nl
$f_1(m)$	& $(1.258 \pm 0.093) \times 10^{-9}$ $\rm{M_{\odot}}$ 	\nl
$T_{\rm transit}$	& $2,450,559.37 \pm 0.54$ HJD				\nl
									\nl
$N$		& $38$							\nl
rms$(O - C)$	& $9.2$ $\rm{m\;s^{-1}}$				\nl
\enddata
\end{deluxetable}

The period and amplitude of the fit are well determined, but because
of the small eccentricity, the longitude $\omega$ and epoch T of
periastron have large uncertainties.  However, the epoch $T_{\rm transit}$
of passage of the companion across the line of sight to the star is
reasonably well determined, as shown in the table.  This value is
included to facilitate possible searches for evidence of a transit
across the disk of the star, should sin $i$ be close to unity.
However, Henry (1997) reports that his 208 photometric data points
over the past four years, when phased at the period in Table 1, make a
transit unlikely.

The derived eccentricity (0.028 $\pm$ 0.040), while close to zero, has
sufficient uncertainty that a value as large as 0.1 cannot be ruled
out.  As noted in  Section 4 below, whether the eccentricity is
significantly different from zero is an important question; more
observations, especially with complete phase coverage, are needed to
pinpoint the value of the eccentricity more accurately.

The dashed line in Figure 1 shows the Keplerian orbital fit from Table
1. Residuals from the fit are shown in the lower part of Figure 1. 
Figure 2 shows the same data phased according to the orbital fit. The
periodic variation of radial velocity is evident.

\begin{figure}[ht]
\plotone{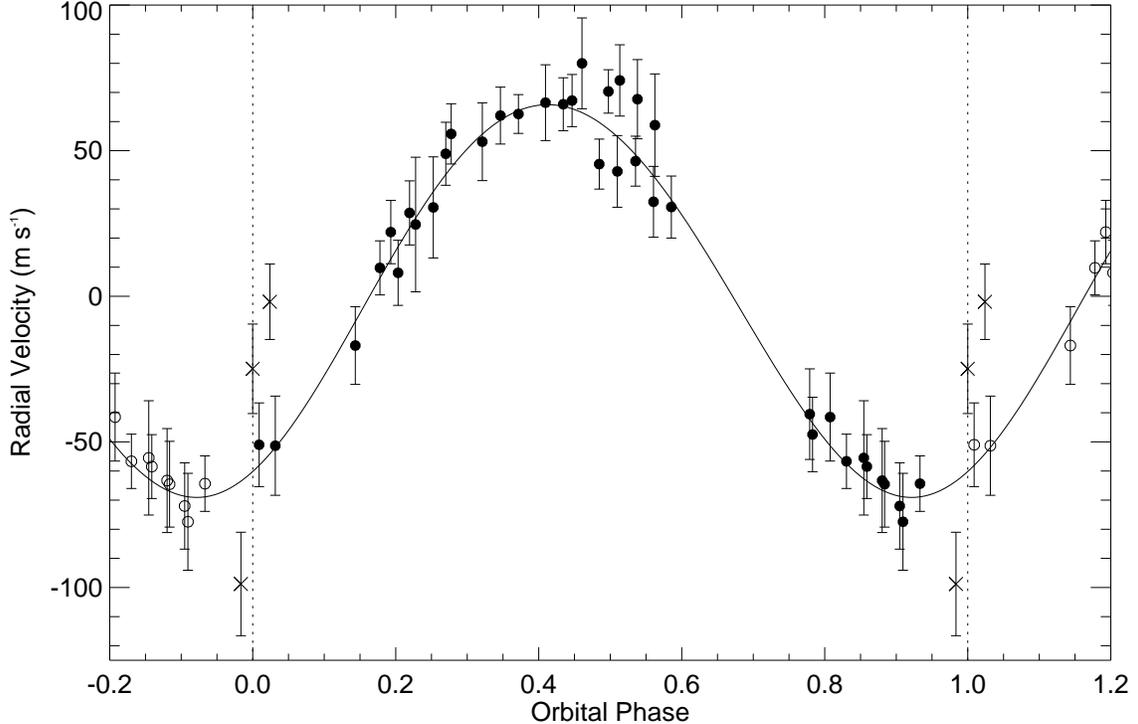}
\caption[]{Radial velocity of $\rho$~CrB (circles), phased according
to the orbital elements listed in Table 1.  Solid line is the orbital
solution. \label{figph}}
\end{figure}

Part of the residuals from the orbital fit may be attributable to
long-term instrumental drifts of uncertain origin.  In Figure 1, such
drifts are suggested especially by the groupings of residuals for the four
1996 observing runs (HJD $<$ 2,450,450); however the smaller rms residuals
(only 5 m s$^{-1}$) of the data points for the 1997 January, February,
and March observing runs indicate that the problem may have largely
disappeared after an instrumental realignment in early January of 1997.
Here we take the conservative approach of not correcting for long-term
drifts in the orbital analysis.

\section{DISCUSSION}

The data plotted in Figure 1 clearly indicate a nearly sinusoidal
variation in the measured radial velocity of $\rho$~CrB, with period
of 39.6 days and amplitude of 67 m s$^{-1}$.  The obvious
interpretation is that $\rho$~CrB is orbited by a low mass companion,
which produces a reflex motion of $\rho$~CrB itself as they revolve
about their common center of mass.  If the mass of $\rho$~CrB is 1.0
$M_{\odot}$ (Section 2), the minimum mass $m_2$~sin$i$ of the
companion is 1.1 Jovian masses, and the orbital radius $a_2$ is 0.23
AU.

Other interpretations could be considered.  It is possible, although
unlikely (Soderblom 1985), that in spite of the approximately 20 day
rotation period implied by the mean Ca II flux level, the actual
rotation period of $\rho$~CrB is 40 days.  In that case, an apparent
periodic variation of radial velocity could in principle be caused by
changes of the photocenter of the rotating star by periodic passage of
a large spot across its disk.  However, an amplitude as
large as we observe would require about 3\% spot coverage, which would
produce a photometric modulation of order 0.03 magnitude.  
Henry (1997) reports an upper limit of 0.0002 $\pm$ 0.0001 magnitude
for any variability at the period of 39.65 days.  This result
effectively rules out the above interpretation.  A rotating magnetic
region with a convective blueshift which somehow produces no
photometric signature could be considered, but the more magnetically
active Sun shows no spurious radial velocity variations as large as
one-tenth the amplitude we observe on $\rho$~CrB (see
\markcite{mcmillan}McMillan et al.\ 1993).  One might postulate a
non-radial pulsation which does not produce any disk-integrated
photometric modulation, as Gray \markcite{gray3}(1997) has suggested
to explain the 4.2-day periodicity observed in the radial velocity of
51 Pegasi \markcite{mayor}(Mayor \& Queloz 1995).  However, the 40 day
period of the radial-velocity variations in $\rho$~CrB is extremely
long compared to the dynamical timescales characterizing
acoustic-gravity modes in Sun-like stars (roughly 1 hour).  Moreover,
the amplitude of 67 m s$^{-1}$ is enormous compared to the 1 m
s$^{-1}$ amplitude of the combined solar ``5-minute'' p mode
pulsations, or of the (so far undetectably small) solar g modes.
Furthermore, the abovementioned stringent limits on photometric
variation are extremely difficult to reconcile with radial or
low-degree nonradial pulsations at an amplitude approaching the
observed velocity amplitude.  For these reasons, we take the view that
the radial velocity variation observed in $\rho$~CrB is best
explained as the result of an orbiting companion.

The companion to $\rho$~CrB may be considered to be a ``planet'',
where planets are taken to be stellar companions having masses
comparable to the mass of the giant planets in our solar system (cf
Marcy et al.\ 1997). In addition, it is interesting to inquire if there are
clues whether this companion was created like the planets
in our own solar system---that is, by agglomeration of planetesimals in
a circularly rotating circumstellar disk (e.g. Lissauer 1995), or
whether it might have been created in the manner of brown
dwarfs---that is, by fragmentation within a collapsing molecular
cloud, as the low-mass tail of a binary star creation.  While it has
been suggested that a lower limit to the mass of brown dwarfs formed
in this way is about 10 $M_{Jup}$ due to opacity effects (e.g. Silk
1977), it has recently been argued \markcite{black}(Black 1997) that
uncertainties in the binary formation process allow for brown dwarf
companions with much lower mass, even overlapping the mass range of
giant planets, so by itself the mass of the companion to $\rho$~CrB
does not determine its mode of origin.

Binary stars, and by extension stars with brown dwarf companions,
typically form with significant eccentricity.  If the orbital
period is relatively short, the orbit may be circularized by tidal
effects.  However, main-sequence binaries with periods as long as the
40 day period of $\rho$~CrB generally do not have circular orbits.
For example, \markcite{latham}Latham et al. (1992), in an analysis of
old (10-15 Gyr) binaries in the halo, showed that only those with
periods less than about 19 days were circularized.  In addition, an
analysis which takes into account the effect of tides on the planet
itself \markcite{rasio2}(Rasio et al. 1996) implies a timescale for
orbital circularization of the $\rho$~CrB system of more than
$10^{12}$ yr.  Thus it appears unlikely that $\rho$~CrB and its
companion formed in the manner of binary stars; formation of the
companion by agglomeration of dust and planetesimals in a dissipative
circumstellar disk around $\rho$~CrB seems more plausible.

The companion to 47 UMa \markcite{butler1}(Butler \& Marcy 1996) is
similar to the companion to $\rho$~CrB, in having both near-zero
eccentricity and an orbital period long enough that tidal
circularization should not have occured over the lifetime of the
system. However, the companion to $\rho$~CrB orbits 9 times closer to
its parent star than does the companion to 47 UMa (at 2.1 AU), and 23
times closer than does Jupiter (at 5.2 AU).  The companions to both 47
UMa and $\rho$~CrB lie inside the radius of the ice condensation zone,
which has been argued \markcite{boss}(Boss 1995) to be the minimum
radius for formation of giant planets.  The existence of a giant
planet, in near-circular orbit as close as 0.23 AU to its parent star
but still outside the tidal circularization radius, sharpens the
question of whether such bodies could be formed {\it in situ}, or if
not, how they could migrate inward and stop in a stable orbit at their
present location.

\acknowledgements
We are grateful to the Mt. Hopkins observing and support staff, especially
Wayne Peters, Bastiaan van't Sant, Perry Berlind, and Jim Peters, for help
with the observations.  We thank Paul Butler and Geoffrey Marcy for obtaining
very high resolution spectra of our I$_2$ absorption cell using the
McMath-Pierce FTS spectrograph.  We are indebted to Guillermo Torres and David
Latham for useful discussions.  We thank Gregory Henry for allowing us to
quote his photometric results on $\rho$~CrB in advance of publication.  We are
grateful to the referee, Geoffrey Marcy, for careful reading of the manuscript
and helpful suggestions. In preparation of this paper, we made use of the
Simbad database operated at CDS, Strasbourg, France and the NASA Astrophysics
Data System.


\begin{references}
\reference{alexander}Alexander, J.\ B.\ 1967, \mnras, 137, 41

\reference{alonso}Alonso, A., Arribas, S., \& Martinez-Roger, C.\
1996, A\&AS, 117, 227

\reference{argue}Argue, A.\ N.\ 1963, \mnras, 125, 557

\reference{baliunas1}Baliunas, S. L, Donahue, R. A., Soon, W. H. et
al. 1995, \apj, 438, 269.

\reference{baliunas2}Baliunas, S. L., Sokoloff, D., \& Soon, W. H. 1996,
\apjl, 457, L99

\reference{black}Black, D. C.\ 1997, preprint

\reference{boss}Boss, A. P. 1995, Science, 267, 360

\reference{bressan}Bressan, A., Fagotto, F., Bertelli, G., \& Chiosi,
C.\  1993, \aap, 100, 647

\reference{brown}Brown, T. M., Noyes, R.W., Nisenson, P., Korzennik, S., \& Horner, S.\ 1994, \pasp, 106, 1285

\reference{butler1}Butler, R. P., \& Marcy, G. W. 1996, \apjl, 464, L153

\reference{butler2}Butler, R. P., Marcy, G. W., Williams, E., McCarthy, C., 
Dosanjh, P., \& Vogt, S. S. 1996, \pasp, 108, 500

\reference{cayrel}Cayrel de Strobel, G.\ 1996, A\&AR, 7, 243

\reference{charbonneau}Charbonneau, P.\ 1995, \apjs, 101, 309

\reference{gliese}Gliese, W.\ \& Jahreiss, H.\ 1991,
Preliminary Version of the Third Catalogue of Nearby Stars.

\reference{gratton}Gratton, R.\ G., Carretta, E., \& Castelli, F.\ 1996, \aap, 
314, 191

\reference{gray1}Gray, D.\ F.\ 1992, \pasp, 104, 1035 

\reference{gray2}Gray, D.\ F.\ 1994, \pasp, 106, 1248

\reference{gray3}Gray, D.\ F.\ 1997, \nat, 385, 449

\reference{henry}Henry, G.\ 1997, personal communication

\reference{hernshaw}Hernshaw, J.\ B.\ 1974, \aap, 36, 191

\reference{horner}Horner, S. 1996, \apj, 460, 449

\reference{jaschek}Jaschek, C., Conde, H., \& de Sierra, A.\ C.\  1964, 
Catalogue of Stellar Spectra Classified in the Morgan-Keenan System, 
Observatorio Astronomico de la Universidad Nacional de la La Plata, 
La Plata

\reference{kunzli}K\"unzl, M., North, P., Kurucz, R.\ L., \& Nicolet,
B.\ 1997, A\&AS, 122, 51

\reference{latham}Latham, D. W., Mazeh, T., Torres, G., Carney, B. W., 
Stefanik, R. P., \& Davis, R. J. 1992, in Binaries as Traces of Stellar 
Formation, ed. A. Duquennoy, \& M. Mayor (Cambridge: Cambridge Univ. 
Press), 278

\reference{lissauer}Lissauer, J. J.\ 1995, Icarus, 114, 217

\reference{marcy}Marcy, G. W., Butler, R. P., Williams, E., Bildsten, L.,
Graham, J. R., Ghezx, A. M., \& Jernigan, J. G.\ 1997, ApJ, in press
(to appear in June issue, Vol. 181)


\reference{mayor}Mayor, M., \& Queloz, D.\ 1995, \nat, 378, 355

\reference{mcmillan}McMillan, R. S., Moore, T. L., Perry, M. L., \&
Smith, P. H.\ 1993, \apj, 403, 801

\reference{naur}Naur, P.\ 1955, \apj, 122, 182

\reference{noyes1}Noyes, R.\ W., Hartmann, L.\ W., Baliunas, S.\ L., Duncan, 
D.\ K., \& Vaughan, A.\ H.\ 1984, \apj, 279, 763

\reference{noyes2}Noyes, R.\ W., Jha, S., Korzennik, S., Krockenberger, M., 
Nisenson, P., Brown, T., Kennelly, E., \& Horner, S. 1997, in ``Planets Beyond
the Solar System and the Next Generation of Space Missions", Space
Telescope Science Institute Workshop, \pasp, in press

\reference{rasio2}Rasio, F. A., Tout, C. A., Lubow, S. H., \& Livio, M. 1996, 
\apj, 470, 1187

\reference{roman1}Roman, N.\ G.\ 1952, \apj, 116, 140

\reference{roman2}Roman, N.\ G.\ 1955, \apjs, 2, 195

\reference{rufener}Rufener, F.\ 1976, A\&AS, 26, 275

\reference{silk}Silk, J. 1977, \apj, 214, 152 

\reference{soderblom}Soderblom, D.\ 1985, \aj, 90, 2103

\reference{soon}Soon, W.\ 1997, personal communication

\reference{taylor}Taylor, B. J.\ 1994, \pasp, 106, 144

\reference{wallerstein}Wallerstein, G.\ 1962, \apjs, 6, 407

\end{references}
\end{document}